# Label-free optical vibrational spectroscopy to detect the metabolic state of *M. tuberculosis* cells at the site of disease

Vincent O. Baron[1, ‡], Mingzhou Chen[2, ‡, *], Simon O. Clark[3], Ann Williams[3], Robert J. H. Hammond[1], Kishan Dholakia[2] and Stephen H. Gillespie[1, *]

[1] School of Medicine, University of St Andrews, St Andrews, UK, KY16 9TF
[2] SUPA, School of Physics and Astronomy, University of St Andrews, KY16 9SS, St Andrews, UK
[3] Public Health England, Porton Down, Salisbury, Wiltshire, SP4 0JG, UK
[‡] These authors contributed equally to this work
*Corresponding authors: mingzhou.chen@st-andrews.ac.uk or shg3@st-andrews.ac.uk

## Abstract

Tuberculosis relapse is a barrier to shorter treatment. It is thought that lipid rich cells, phenotypically resistant to antibiotics, may play a major role. Most studies investigating relapse use sputum samples although tissue bacteria may play an important role. We developed a non-destructive, label-free technique combining wavelength modulated Raman (WMR) spectroscopy and fluorescence detection (Nile Red staining) to interrogate *Mycobacterium tuberculosis* cell state. This approach could differentiate single "dormant" (lipid rich, LR) and "non-dormant" (lipid poor, LP) cells with high sensitivity and specificity. We applied this to experimentally infected guinea pig lung sections and were able to distinguish both cell types showing that the LR phenotype dominates in infected tissue. Both *in-vitro* and *ex-vivo* spectra correlated well, showing for the first time that *Mycobacterium tuberculosis,* likely to be phenotypically resistant to antibiotics, are present in large numbers in tissue. This is an important step in understanding the pathology of relapse supporting the idea that they may be caused by *M. tuberculosis* cells with lipid inclusions.

## Introduction

Tuberculosis is now established as the most important cause of death due to infectious disease, yet treatment has not improved in fifty years. Relapse after successful treatment is the major barrier to shorter therapy for tuberculosis as has been confirmed by recent tuberculosis clinical trials where more bactericidal regimens have failed due to higher relapse rate [1-3]. Despite its importance, we know very little about the bacteriology of relapse even though this outcome is possible with patients who clear their sputum rapidly [4]. We need improved and non-destructive methods to study mycobacteria at the site of the disease.

It is often speculated that relapse is linked to bacteria that survive treatment. Several studies demonstrate the accumulation of lipids in intracellular bodies that are associated with a lower metabolic rate [5-8]. It has been shown recently that these lipid body positive cells are up to 40 times more resistant to key components of the treatment regimen such as isoniazid [9].

An attractive approach to investigate the lipid content of *Mycobacterium tuberculosis* would be to use an all-optical label-free method to examine bacterial cells in tissue in a way that would, for example, allow additional immunology studies. Optical interrogation can lead to inelastic scattering of light and distinct vibrational bands in subsequent spectra that can be used to distinguish the molecular content of the bacteria under investigation. In particular, Raman spectroscopy has been used previously as a means of identifying bacterial taxonomy at a single cell level in a range of species including mycobacteria [10,11]. These studies, however, have only previously been performed on isolated cultured cells [12] and isolated cells from sputum [13].

This letter reports, for the first time, the metabolic cell state of mycobacteria by exploring the lipid content of individual cells in tissue using wavelength modulated Raman (WMR) spectroscopy. WMR analysis uses a scan of the laser wavelengths, rather than a single wavelength, for Raman excitation combined with subsequent multivariate statistical analysis that removes all background fluorescence [14,15]. WMR spectroscopy shows an increase in signal-to-noise ratio compared to five methods including standard Raman spectroscopy [16]. Using this approach, we were able to detect LR *M. tuberculosis* cells in infected tissue with very high sensitivity and specificity. This is a major step towards understanding the pathogenesis of tuberculosis at the infection site and provides a paradigm for its application to a broad range of infectious diseases.

## Results
### Single cell discrimination between lipid rich and lipid poor bacteria

We separated mixed cultures of *M. smegmatis, M. bovis* (Bacillus Calmette-Guérin, BCG) and *M. tuberculosis* into lipid rich (LR) and lipid poor (LP) fractions with greater than 90% purity using our previously published method [9], and performed WMR spectroscopy on the separated cells.

We recorded Raman spectra from ~60 individual lipid rich (LR) and lipid poor (LP) mycobacteria cells (*M. smegmatis,* BCG and *M. tuberculosis*) *in-vitro*. The spectra from the mycobacterial species are illustrated in the **Supplementary Section 1**. In wavelength modulated Raman spectra (WMR spectra) zero-crossings are equivalent to peak positions in standard Raman spectra and the peak-to-valley corresponds to the peak intensity in standard Raman spectra. The two phenotypes mainly differ in two lipid peaks at 1300 cm$^{-1}$ (designated lipid band A) and at 1440-1450 cm$^{-1}$ (designated lipid band B), see **Supplementary Fig. 1** (see **Supplementary Section 4** for the peak assignment). LR cells showed higher Raman peak intensity in both lipid bands (A and B) compared to LP cells. The two phenotypes are distinguished with high specificity and sensitivity, *M. smegmatis* (93.8%/96.8%), *BCG* (100%/96.8%) and *M. tuberculosis* (92.6%/96.1%) (see **Supplementary Fig. 1**).

**Identification of *M.tuberculosis* in unstained infected lung tissue sections**

Having created a tool to determine the lipid status of mycobacterial cells non-destructively we applied the technique to unstained, formaldehyde-fixed guinea pig lung tissue infected by *M. tuberculosis* to understand the cell state of the bacteria at the site of disease (see **Methods section** for more details). Guinea pig tissue was used in the analyses because the lesions which develop during pulmonary tuberculosis in guinea pigs are histologically similar to the disease seen in humans [17]. In total, 107 single *M. tuberculosis* cells were interrogated by WMR spectroscopy from the lung alveoli (**Fig. 1c** and **Fig. 2b**). An example of a single bacillus in an alveoli is shown in **Fig. 1a**. The tissue section is attached to a quartz coverslip. The background of the lung alveoli was compared to the signal from a clean quartz slide. The two signals were found very comparable and no important signal is coming from the lung alveoli itself (**Fig. 1b**). The **Fig. 1c** shows the average WMR spectrum of the bacteria acquired in the alveoli with the two background signals.

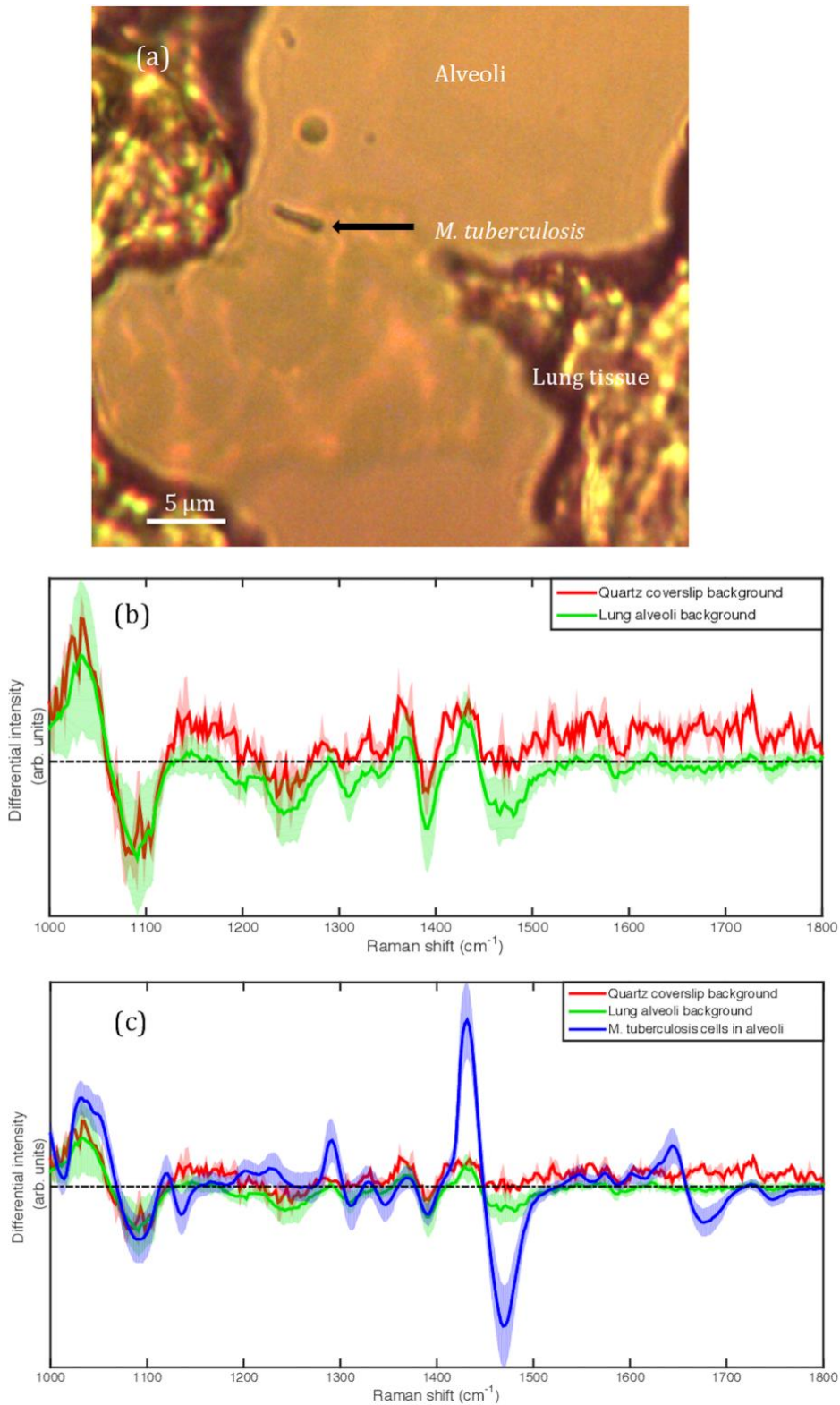

***Figure 1*** *investigation of the impact of the lung alveoli to the bacterial Raman spectrum. The **Fig. 1a** shows an example of single bacillus in a lung alveoli interrogated by WMR spectroscopy. The scale bar corresponds to 5 µm. The **Fig. 1b** presents the average Raman spectrum of the quartz slide surface (red line) and the average Raman spectrum of the lung alveoli (background)(green line). The tissue section like any other preparation interrogated with WMR spectroscopy is on a quartz coverslip. In the **Fig. 1c** the average spectrum of M. tuberculosis cells acquired in tissue are added (blue line) and compared with the lung alveoli and the quartz slide background. In both **Fig. 1b** and **Fig. 1c** the shaded coloured areas represent the standard deviation. In **Fig. 1b** and **Fig. 1c** the x-axis is in wavenumbers the y-axis represents the differential intensity in arbitrary units.*

It was not possible to relate our *in-vitro* and *ex-vivo* data directly as it was noted that lipid bands A and B were impacted differently by formalin fixation and freezing [11,18] (**Fig. 2a and 2b**). In *M. tuberculosis* lipid band A, intensity is much lower when acquired in tissue as compared to *in-vitro*. In contrast, there is less variation in lipid band B intensity between *in-vitro* and *ex-vivo* samples. Lipid band B was therefore used in subsequent examinations (**Fig 2c**). To compare *in-vitro* with *ex-vivo* data we needed to establish a ratio between a band that varies and one that varies very little. Moreover, when *in-vitro* data are compared (**Fig. 2a**) similar peak intensities are found in the band between 1050 cm$^{-1}$ and 1070 cm$^{-1}$ (see **Supplementary Section 4, peak assignments**) for both *M. tuberculosis* LR and LP phenotypes. This was also true for *M. tuberculosis* in *ex-vivo* samples. Thus, this band is ideal as an internal reference and we, therefore calculated the peak-to-peak ratio ($R_{band\ B/Ref\ band}$) between the maximum peak intensity of lipid band B and the maximum peak intensity in the internal reference band (1050 cm$^{-1}$ and 1070 cm$^{-1}$) for each WMR spectrum acquired from *M.tuberculosis* in lung tissue (**Fig. 2b**) and also for each WMR spectrum acquired from *M.tuberculosis in-vitro* (**Fig. 2a**). A higher ratio indicates a higher lipid concentration in the cell.

Using this technique it is notable that the distribution of values for LR and LP cells *in-vitro* overlaps, but the peak of the distributions are clearly separated (**Fig. 2c**). We are able to observe a lipid ratio value $R_{band\ B/Ref\ band}$ analogous to both LP and LR among cells in tissue. The $R_{band\ B/Ref\ band}$ obtained from WMR spectra of cells in infected lung tissue shows a wide distribution confirming the presence of both LR and LP phenotypes. It is notable that the distribution of cell state in tissue is skewed towards LR phenotype (**Fig. 2c**). A similar result was obtained when ($R_{band\ A/Ref\ band}$), with band A adjusted for the losses associated with fixation and freezing was used (see **Supplementary Section 3**).

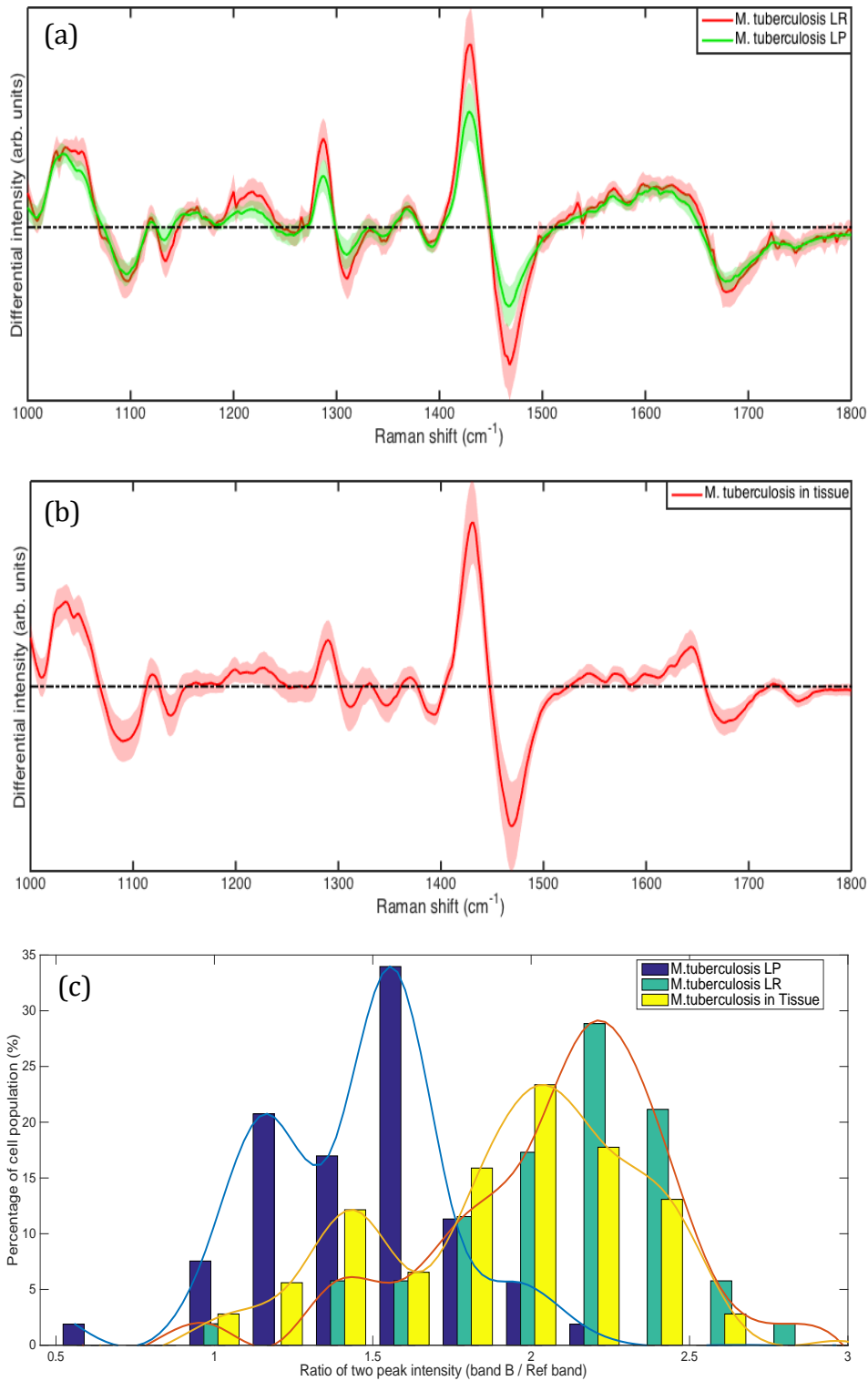

***Figure 2 (a)*** *WMR spectra of LR and LP M. tuberculosis cells. Mean spectra of both LR (red curves) and LP (green curves) cells are calculated from ~60 WMR spectra are taken from single bacteria cells. The shaded area represents one standard deviation. The x-coordinate corresponds to the Raman shift (in wavenumber, cm$^{-1}$) and the y-coordinate the differential Raman intensity in arbitrary units. **(b)** WMR spectra from M. tuberculosis cells in the infected lung tissue. The solid red line shows the mean WMR spectrum averaged over from all spectra of 107 cells taken at the single bacterium cell level. The colour-shaded area represents the associated single standard deviation.*

*The acquisition time for each single bacterium was 150 seconds in total (see **Methods section** for more details). **(c)** Percentage of M. tuberculosis in-vitro LR, LP and form tissue with a given peak-to-peak ratio (R $_{band\ B/Ref\ band}$) calculated by dividing the lipid band B intensity value by an internal reference band (Ref band: 1050 cm$^{-1}$ to 1070 cm$^{-1}$) peak intensity for each in-vitro M. tuberculosis WMR spectra (**Fig. 2a**) and each M. tuberculosis acquired in the lung tissue (**Fig. 2b**). The x-coordinate corresponds to R $_{band\ B/Ref\ band}$ between 0.5 and 3 and the y-coordinates represent the percentage of the bacterial population for each specific lipid ratio value. The blue, green and yellow bars correspond to the in-vitro LP, the in-vitro LR and the ex-vivo populations respectively. The lines in **Fig. 2c** are included to illustrate the shape of the distribution.*

**Identification of LR and LP single bacterium in the *M.tuberculosis* infected tissue**

To ensure that the match between *in-vitro* and *ex-vivo* cells that we demonstrate is not an *in-vitro* artefact, we obtained an alternative measure of lipid cell state from tissue in parallel with WMR spectroscopy. To do this we used our previously published staining method [9]. As shown in **Fig. 3**, all LP cells appear only red caused by excitation from cell wall polar lipids. LR bacteria are red from polar cell wall lipids and are distinguished from LP by green fluorescence emission caused by the non-polar lipid body. We studied *in-vitro* stained cells and stained cells in guinea pig infected lung. Nile red staining does not impact significantly the WMR spectrum of bacteria as shown in **Supplementary Section 6.** We scored the Nile Red stained bacilli as either LR or LP as previously described and recorded the Raman spectrum from these defined cells (**Fig. 4**). This showed that, for both *in-vitro* cells and cells in tissue that the main source of variability is found in bands A and B. We also showed that *in-vitro* LR and LP cells could be discriminated by WMR spectroscopy (sensitivity (84.0%) and specificity (80.2%)) (**Fig. 4a and 4b**). LP cells cluster closely but LR are more dispersed. The presence of green fluorescence emission coming from intracellular lipid bodies correlates with higher Raman lipid peaks (in both lipid band A and B).

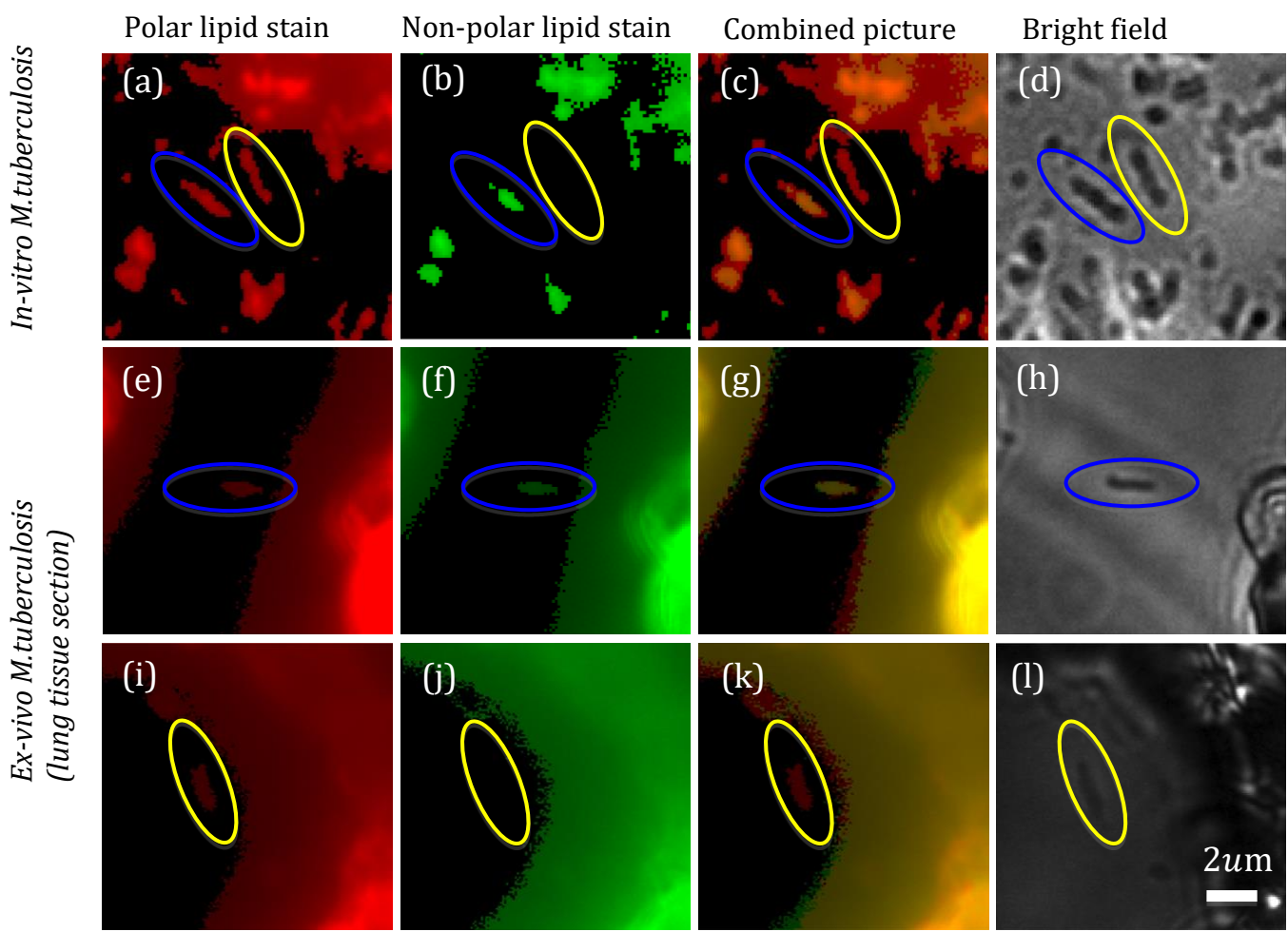

*Figure 3* M. tuberculosis in-vitro and from frozen lung tissue section stained with Nile red and observed on the fluorescence-Raman combined system (more details in the method section). *(a, b, c, d)* show in-vitro M. tuberculosis and *(e, f, g, h, i, j, k and l)* show M. tuberculosis in lung tissue. *(a, e, i)* show the red fluorescence (polar lipids) and *(b, f, j)* show the green fluorescence emission (non-polar lipids). *(c, g, k)* represent the pseudo images made by combining red and green fluorescence emissions images. Bright field images (White light) are shown in **d, h and l**. Both red and green fluorescence emissions can be observed in LR cells encircled in blue, while only red fluorescence emission is presented in LP cells encircled in yellow. The size of bacteria can be calculated from the scale bar shown in **l** (typically between 0.5 – 1μm in width and 2 - 4μm in length).

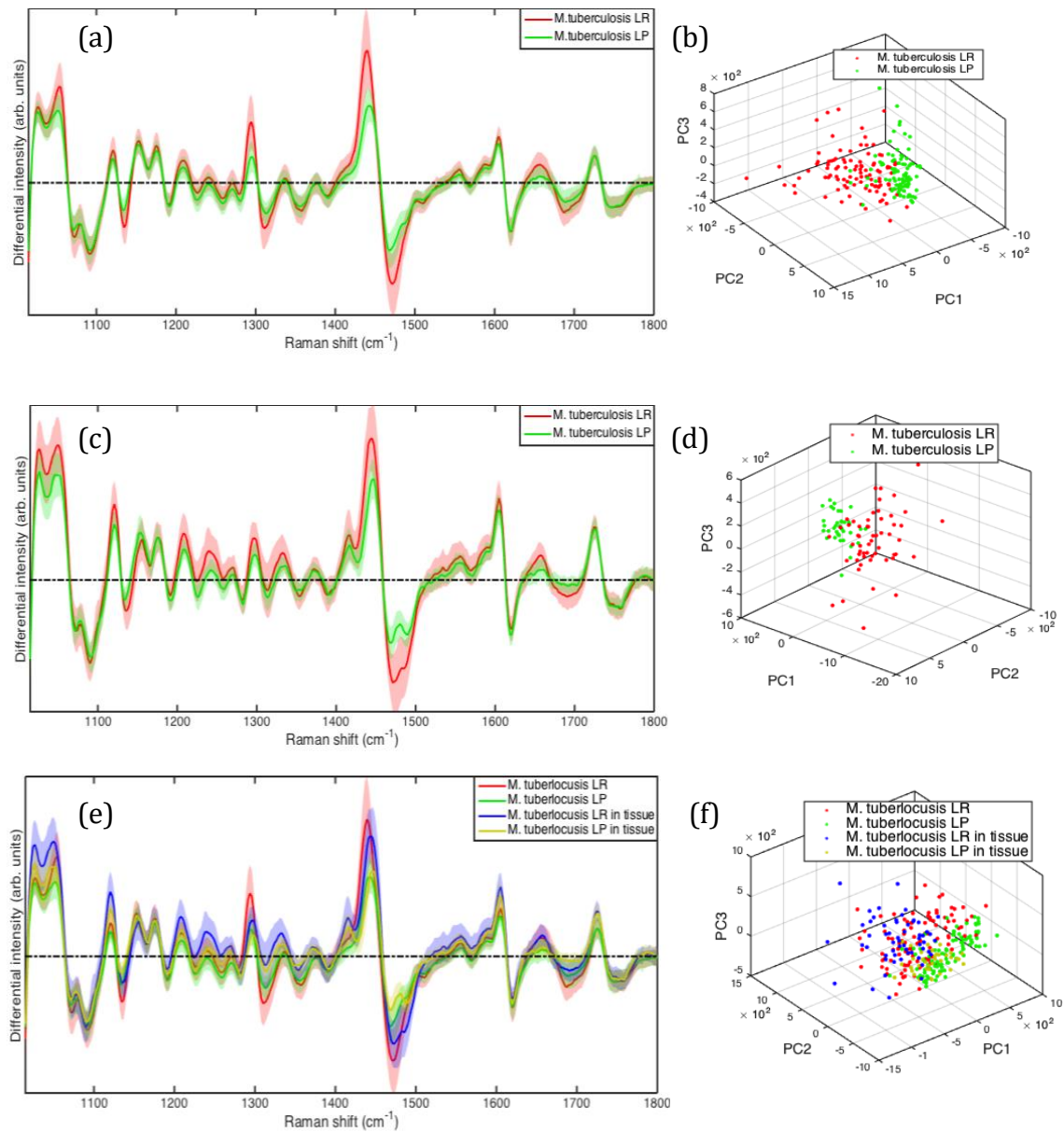

*Figure 4* M.tuberculosis in-vitro and from infected lung tissue WMR spectra and PCA clusters. LR and LP bacteria are pre-assigned by fluorescence signals from Nile red stained bacteria. (**a**) Mean spectra of both LR (red curve) and LP (green curve) cells averaged over 100 WMR spectra taken in-vitro. The light coloured area represents the associated single standard deviation. The first three PCs for both LR and LP cells form two 3D clusters in (**b**). The LR cluster shows a broader cluster compared to LP group. (**c, d**) M. tuberculosis from lung tissue WMR spectra and PCA clusters. (**c**) Mean spectra of both LR (red curve) and LP (green curve) cells. The first three PCs for both LR and LP cells form two 3D clusters in (**d**). The LR cluster shows a broader cluster compared to LP group similarly than in-vitro. (**e**) Comparison of in-vitro and lung tissue M.tuberculosis LR and LP cells. In-vitro LR (red) and in-vitro LP (green) were compared to the ex-vivo LR (blue) and ex-vivo LP (yellow) bacilli acquired in infected lung tissue. (**e**) Mean spectra of both lipid rich and lipid poor cells acquired with WMR spectroscopy. (**f**) 3D clusters are plotted with the first three PCs for LR and LP in-vitro cells and the cells from infected tissue.

We then applied this method to *M. tuberculosis* cells (93 in total) in infected lung tissue sections. 56 cells were classified LR by Nile Red staining and 37 as LP (**Fig. 3*e - l***). Using PCA LR and LP cells clustered together although there is a wider distribution for the LR cells as had been noted *in-vitro* (**Fig. 4d**). As in the case of our *in-vitro* data, green fluorescence correlates with higher lipid peaks in lipid bands A and B. A broader cluster in LR group observed in **Fig. 4d** also shows higher variability among the LR population than the LP population in tissue. Thus, it was possible to discriminate between LR and LP cells with high sensitivity (92.6%) and specificity (84.6%) (**Fig. 4c and 4d**). This is illustrated by combing the figures to demonstrate the overlap between the differing cell types *in-vitro* and *ex-vivo* which cluster together (**Fig. 4e and 4f**). Using this measure, we are able to demonstrate that *M. tuberculosis* cells in the lung are predominately LR (60%).

## Discussion

Preventing relapse is the main reason for prolonged treatment in tuberculosis, yet the causes are still not fully understood [1,4]. In particular we have shown that some patients who become sputum culture negative rapidly still go on to relapse [4]. Our research provides a new tool to unravel this problem. This is the first time the label-free vibrational spectroscopy has been used to study the pathogenesis of tuberculosis in situ. Whilst Raman has been used previously to distinguish bacterial genus and species [10,11], our work makes a major leap forward giving us the ability to distinguish bacterial cell state at the site of disease. Reducing the Raman signal window size may affect the results slightly but does so within a small standard deviation of 0.044 in the obtained sensitivity and specificity (see in **Supplementary Section 2** for more details). The possibility of this window reduction with a similar performance actually gives us a way to make our system more compact and to have higher resolution, i.e. using a grating only over a range of 1400-1500cm$^{-1}$ and a shorter acquisition time. Future developments of our work therefore could lead to wide field video-rate acquisition *in-vivo* with non-linear vibrational spectroscopy approaches using suitable tuneable lasers for coherent anti-Stokes Raman scattering or stimulated Raman scattering [19,20]. This confirms that vibrational spectroscopy can be a powerful methodology that can be applied to pathogenesis and pharmacodynamics studies.

The study of the physiological status of *M. tuberculosis* bacteria has been limited to the sputum while may not fully represent the cells responsible for relapse. The physiological state of the bacteria that never get into the sputum are missed by such approaches [21]. Using WMRS we have been able to differentiate lipid rich from lipid poor mycobacteria in tissue and consequently our data bridges the gap between tissue and sputum. Importantly, current methods, for example Nile Red staining (see in **Supplementary Section 5** for more details), require destructive techniques, which means that no additional studies can be performed [22]. The preparations that we have used could be re-examined with, for example, immunohistochemical methods to differentiate macrophage activation near to *M. tuberculosis* for which the metabolic state is now known. This opens up a wide range of new research possibilities that we will explore in future experiments.

The distinction between LR and LP is important as the LR phenotype has been associated with a reduced metabolic activity and down regulation of a series of enzymes [6]. More importantly, such LR cells have been shown to be phenotypically resistant to anti-tuberculosis antibiotics [7,23]. Antibiotic susceptibility testing of purified populations of LR and LP cells suggest that the resistance can increase from between 3-5 times (ciprofloxacin) to more than 20 times for isoniazid and rifampicin [9]. We have now shown that in tissue the majority of the cells that we examined had a LR phenotype that implies a significant degree of phenotypic resistance. Our confidence in this result is increased by the concordance of *in-vitro* and *ex-vivo* clustering as demonstrated in **Fig. 4f**; confirming as well that no significant signal from the lung is participating to the WMR spectra (**Fig 1**). Thus, our observation may throw some light on the mismatch between standard anti-tuberculosis susceptibility testing and the outcome of treatment in patients and the overall result of clinical trials. Many patients relapse after apparently successful treatment and this is thought to be due to dormant bacteria able to survive treatment and regrow [4]. The presence of lipid bodies positive bacteria in TB infected patients' sputum has been linked to a higher risk of poor outcome [24]. However sputum data might not be representative of the bacterial population living in patient lungs. Our data derived from lung tissue supports the idea that relapses may be caused by LR *M. tuberculosis* with phenotypic resistance to the prescribed treatment. It should be noted, however, that we characterised bacteria in the alveoli and not in the complex structure within solid tissue. Future developments will enable us to achieve this. Thus, using Raman spectroscopy we have shown that lipid rich bacteria are present in the lung at the site of disease and are the predominant cell type (**Fig. 2c, Fig. 4c and 4d**). This observation suggests that this characteristic may allow LR cells to survive antibiotic chemotherapy.

# Methods

## Bacterial culture conditions

*M. bovis* (BCG, NCTC 5692), *M. smegmatis* (NCTC 8159), *M. tuberculosis* (NCTC7416) were grown at 37°C in Middlebrook 7H9 medium (Sigma-Aldrich) supplemented with 4 ml of 50% glycerol (for 450 ml) (Sigma-Aldrich) and 0.05% tween80 (Fisher BioReagents). BCG and *M. tuberculosis* culture medium were also supplemented with 1 vial of Middlebrook ADC enrichment (Sigma-Aldrich).

## Animal infection

The lung sample analysed in these studies were from specific pathogen-free guinea pigs infected with an aerosol dose of 10–50 CFU (retained dose in the lung) of *M. tuberculosis* H37Rv (NCTC cat. no. 7416) [25]. Nose-only aerosol challenge was performed using a fully contained Henderson apparatus as previously described [26] and [27] in conjunction with the AeroMP (Biaera) control unit [28].

Guinea pig experimental work was conducted according to UK Home Office legislation for animal experimentation and was approved by the UK Home Office. All animals were weighed weekly and observed daily in order to monitor any adverse effects.

Four weeks post-challenge, animals were killed by an overdose of sodium pentobarbital. At necropsy, the lung was immediately excised and sections of the left and right cranial lobes and right caudal lobes were fixed in 10% neutral buffered formalin for up to 18 months.

## Frozen section

Following fixation, the lung tissue was placed into OCT (optimal cutting temperature) solution (30% sucrose in PBS) to embed and freeze the tissue on dry ice. The fixed and frozen tissue was sectioned into 5μm slices using a cryostat by Amsbio Ltd, UK (AMS Biotechnology (Europe) Limited / Registered office: 184 Park Drive, Milton Park, Abingdon OX14 4SE / Registered in England & Wales: Company number 2117791 / ISO 9001:2008 registered firm). In this study three serial lung sections coming from one guinea pig were interrogated.

## Nile Red staining of Bacterial culture

The bacteria were stained using a Nile Red stock solution at 250 μg/ml diluted in DMSO (Sigma-Aldrich) and a 1 μl aliquot added to 100 μl of bacterial suspension, vortexed and left for ten minutes in the dark at room temperature. The tubes were spun down at 20,000 g for 3 minutes the supernatant was discarded. The bacteria were washed twice using PBS. The bacterial pellet was resuspended in 20 μl of PBS and 10 μl heat fixed on a glass slide for light microscope or a quartz slide for Raman spectroscopy.

**Nile Red staining of guinea pig lung tissue section**
A 5μm thick guinea pig frozen lung section mounted on quartz coverslip (SPI Supplies, 01015T-AB) was stained using 2 μl of Nile red diluted in PBS, 25 μg/ml final concentration. The 2 μl were placed on the tissue section and the coverslip was then placed on top of a thick quartz slide (SPI Supplies, 01016-AB). The mount was sealed as noted previously. The sample was then interrogated by WMR spectroscopy.

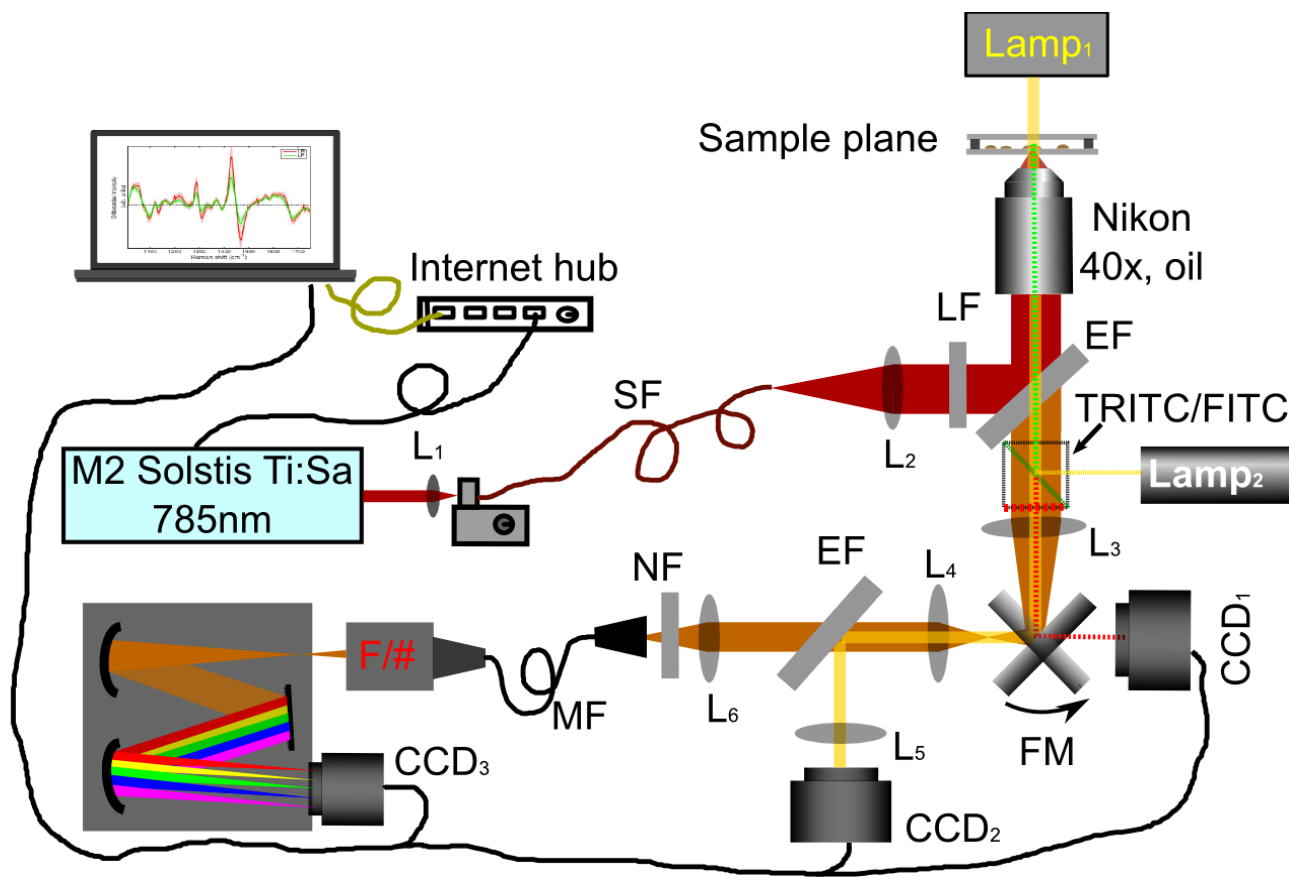

*Figure 5. System setup for a combined Raman-fluorescence microscope:* ***SF****: single-mode fibre;* ***MF****: multi-mode fibre (200um core diameter);* ***LF****: Laser line filter (Semrock LL01-785);* ***EF****: Edge filter (Semrock LPD02-785RU);* ***NF****: Notch filter (Semrock NF03-785E);* ***TRITC/FITC****: Fluorescence Filter cubes (FITC: excitation 475-490nm/emission 500-540nm, TRITC: excitation 545-565nm/emission 580-620nm);* ***FM****: flip mirror;* ***CCD$_1$****: Hamamatsu ORCA_ER;* ***CCD$_2$****:Imaging Source USB camera*

*(DFK 42AUC03);* ***CCD$_3$****:Andor Newton Camera (cooled at -70 °C);* ***L1~L6****: lenses;* ***F/#****: F number matcher.*

**Combined Raman-fluorescence spectroscopy setup**

WMR spectra were recorded by a combined confocal Raman-fluorescence imaging system based on a Nikon microscope (Nikon TE2000-E). A tunable Ti:Sa laser (SolsTis M Squared lasers, 1W@785nm) was focused by a microscope objective (Nikon Plano, 40x, oil) onto a single cell using the arrangement illustrated in **Figure 5**. Excited Raman photons were then collected by a spectrometer formed with a monochromator (Andor Shamrock SR303i, 400 lines/mm grating @850nm) and a cooled CCD camera (Andor Newton, CCD$_3$). With this configuration, the confocal diameter and depth of the system are 5μm and 4.68μm respectively. In order to get enough strong Raman signal from a single cell, we apply a laser dosage at 150mW onto the sample plane. Over a long period of exposure time, this focused laser power didn't show any damage to the cells.

The system is able to switch from Raman spectroscopy to fluorescence imaging by placing FITC/TRITC cubes into the microscope and flipping the flip mirror (FM) to CCD$_1$ (Hamamatsu ORCA-ER). Fluorescence images were taken via standard Nikon fluorescence cubes (FITC and TRITC) using a Nikon fluorescence white light source (Lamp$_2$). Red fluorescence emission (580~620nm) of Nile red stained samples excited by a green wavelength (545~565nm) light shows the signals from the polar lipids whereas green fluorescence emission (500~540nm) excited by a blue wavelength (475~490nm) light shows the signals from the non-polar lipids (lipid bodies).

**WMR spectra**

In order to obtain the WMR spectrum from a single cell, five spectra were recorded continuously over 2.5 minutes with a 30 seconds integration time for each spectrum. During acquisition, the laser line was tuned continuously over a total modulation range of Δλ=1nm. A single WMR spectrum can be produced from these five spectra with all background fluorescence being removed essentially. In the WMR spectrum, all Raman peaks will locate at the zero crossings while their peak intensity will be reflected by the peak-to-valley value around the zero crossing.

**Raman calibration and spectra processing**

Raman spectra were taken from polystyrene beads (1um in diameter) as a control to calibrate the laser line and the drift in the system. The known Raman peak position (1001.4 cm$^{-1}$) of polystyrene bead was used for calculate the laser line. In this way, the drift in the laser line can be monitored during the whole data acquisition procedure. Suppose the drift is very slow (typically <0.2nm over a day), we can calibrate the laser line used for each Raman spectrum from cells through an interpolation. Each Raman spectrum was also normalized by its total intensity (i.e. the integration over the area covered by the spectrum) in order to avoid any fluctuation in the laser power during wavelength modulation. The spectral region between 1000 cm$^{-1}$ and 1800 cm$^{-1}$ was used for the data analysis.

**Principal component analysis (PCA)**

WMR spectra were collected from 60~100 cells from each cell phenotype. PCA was then applied to these training dataset in order to reduce the dimensionality. To distinguish between each two different cell phenotypes, the first seven principal components were taken into account, as they accounted for more than 70% of variances in these training dataset. This algorithm was written in Matlab codes.

**Leave-one-out cross validation (LOOCV)**

The ability of distinguishing between each two different cell subsets was estimated by the method of leave-one-out cross validation. A multiple-dimensional space was defined by the principle components from the training dataset without one Raman spectrum. This leave-out spectrum was then classified in this space with the nearest neighbor algorithm. With this LOOCV method, correct or incorrect predictions for lipid rich and lipid poor cells in the training dataset were then used to estimate the specificity and sensitivity. Confusion matrix can also be estimated from correct and incorrect predictions for more than three cell subsets in the training dataset. This algorithm was written in Matlab codes.

**Acknowledgements**

This work was supported by the PreDiCT-TB consortium [IMI Joint undertaking grant agreement number 115337, resources of which are composed of financial contribution from the European Union's Seventh Framework Programme (FP7/2007-2013) and EFPIA companies' in kind contribution (www.imi.europa.eu)]. M.C. and K.D. thank the UK Engineering and Physical Sciences Research Council (Grant code EP/J01771X/1) and a European Union FAMOS project (FP7 ICT, 317744) for funding. K.D. acknowledges support from a Royal Society Leverhulme Trust Senior Fellowship and the loan of a laser from M Squared Lasers. This work was supported by the Department of Health, UK. The views expressed in this publication are those of the authors and not necessarily those of the Department of Health. We thank the staff of the Biological Investigations Group at PHE Porton for assistance in conducting the study and thank Dr. Peter Caie and Mrs In Hwa Um from School of Medicine in University of St Andrews for their help with the immuno-staining.

**Author contributions**

V.O.B. and M.C. contributed equally to this work. M.C. designed the optical system. V.O.B. prepared the samples. V.O.B. and M.C. performed the experiments and data analysis. V.O.B., M.C., K.D. and S.H.G. contributed to the development and planning of the project, interpretation and discussion of the data and the writing of the manuscript. S.O.C. and A.W. provided TB in-vivo modelling expertise and in-vivo samples and contributed to the writing of the manuscript. R.J.H.H. provided the mouse tissue infected with BCG and characterized the LP/LR cells using Nile red staining and helped with tissue sectioning and contributed to the writing of methods.

**Additional information**

**Competing financial interest**

The authors declare no competing financial interests.